# A New Model for Calculating the Binding Energy of Lithium Nucleus under the Generalized Yukawa Potential and Hellmann Potential

M. Ghazvini[1], N. Salehi[*2] and A. A. Rajabi[1]

[1]Department of physics, Shahrood University, Shahrood, Iran.

[2]Department of Basic Sciences, Shahrood Branch, Islamic Azad University, Shahrood, Iran.

## Abstract

In this paper, the Schrödinger equation for 6-body system is studied. We solved this equation for lithium nucleus by using supersymmetry method with the specific potentials. These potentials are Yukawa potential, the generalized Yukawa potential and Hellmann potential. The results of our model for all calculations show that the ground state binding energy of Lithium nucleus with these potentials are very close to the ones obtained in experiments.

*Keywords:* Lithium nucleus; Binding energy; Schrödinger equation; Supersymmetry; Yukawa potential; generalized Yukawa potential; Hellmann potential.

*PACS numbers:* 21.10.-k ;Properties of nuclei; nuclear energy levels, 21.10.Dr; Binding energies and masses, 24.10.Cn; Many-body theory, 21.30.x; Nuclear forces.

## 1. Introduction

A small number of exact solutions to the Schrödinger equation were obtained historically in the genesis of quantum mechanics. One of the important taskes of quantum mechanics is to solve the Schrödinger equation with the physical potentials. In the past several decades, many efforts have been produced in the literature to study the stationary Schrödinger equation with the central potentials. Until now, the Schrödinger equation for 2 and 3 particles by using several methods such as NU method, Supersymmetry method (SUSY), and Anzast method with different potential is solved [1-9]. Same work is done for other equations (for example Dirac equation, Klein-Gordon equation, ...) [10]. Now we want to do it for 6-body particles (for instance lithium element). In this work the Supersymmetry method is used. Here 6-body Schrödinger equation with the central potentials is studied.

In sections (2) and (3) we review briefly Yukawa Potential, generalized Yukawa potential and Hellmann Potential. In next section, Supersymmetry Method is reviewed. Then we study an analytical solution to the Schrödinger equation for 6-body system and we report the numerical results. Section (6) includes summery on the paper and conclusions.

## 2. The Yukawa Potential

The Yukawa potential is used in various areas of physics to model singular but short-range interactions [11]. In high energy physics, for example, it is used to model the interaction of hadrons in short range gauge theories where coupling is mediated by the exchange of a massive scalar meson [11, 12]. In atomic and molecular physics, it represents a screened Coulomb potential due to the cloud of electronic charges around the nucleus, which could be treated in the Thomas-Fermi approximation that leads to [13]:

$$V = -\frac{V_0 e^{-ar}}{r} \quad (1)$$

where *a* is the screening parameter and is the strength of the Yukawa potential. This potential is often used to compute bound-state normalizations and energy levels of neutral atoms which have been studied over the past years. Now for getting better physical results, we added a term to this potential. So the Generalized Yukawa potential takes the form of:

$$V = \frac{-V_0 e^{-ar}}{r} + \frac{k}{r^2} \quad (2)$$

where *k* is the strength of the Generalized Yukawa potential.

### 3. The Hellmann potential

The Hellmann potential $V(r)$ given by:

$$V = \frac{V_0 e^{-ar}}{r} - \frac{k}{r}. \quad (3)$$

where $k$ and $V_0$ are the strengths of the Coulomb and the Yukawa potential respectively, and $a$ is the screening parameter. It has many applications in atomic physics and condensed-matter physics [14-24]. The Hellmann potential, with $V_0$ positive, was suggested originally by Hellmann [15, 25] and henceforth called the Hellmann potential if $V_0$ is positive or negative. The Hellmann potential was used as a model for alkali hydride molecules [17]. It was used also to represent the electron-ion [18, 19] and electron core interaction [20, 21]. It has also been shown that the main properties of the effective two-particle interaction for charged particles in polar crystals may be described by this potential [22-24].

### 4. Supersymmetry Method in Quantum Physics

We start noticing that we know the ground state function of a 1dimension problem; we can find also the potential, up to a constant [9]. Taking the ground state energy to be zero, from the TISE we have:

$$H_1 \Psi_0(x) = -\frac{\hbar^2}{2m} \frac{d^2 \Psi}{dx^2} + V_1(x) \Psi_0 = 0 \quad (4)$$

$$V_1 = \frac{\hbar^2}{2m} \frac{\Psi_0''(x)}{\Psi_0(x)} \quad (5)$$

We can try to factorize the Hamiltonian with the ansatz:

$$H_1 = A^+ A = H - E_0 \quad (6)$$

where:

$$A = \frac{\hbar}{\sqrt{2m}} \frac{d}{dx} + W(x), A^+ = -\frac{\hbar}{\sqrt{2m}} \frac{d}{dx} + W(x) \tag{7}$$

and $W(x)$ is called the superpotential and can be written in terms of the ground state as:

$$W(x) = -\frac{\hbar}{\sqrt{2m}} \frac{\Psi'_0}{\sqrt{\Psi_0}} \tag{8}$$

By writing $V_1$ in terms of $W(x)$, we obtain the Ricatti equation:

$$V_1(x) = W^2(x) - \frac{\hbar}{\sqrt{2m}} W'(x) \tag{9}$$

So now we can build up a SUSY theory searching for the SUSY partner Hamiltonian associated to $H_1$, namely $H_2 = AA^+$. This second Hamiltonian corresponds to a new potential:

$$H_2 = -\frac{\hbar^2}{2m} \frac{d^2}{dx^2} + V_2(x), V_2(x) = W^2(x) + \frac{\hbar}{\sqrt{2m}} W'(x) \tag{10}$$

### 5. The Exact Solution of The Schrödinger Equation for Yukawa Potential, Generalized Yukawa Potential and Hellmann Potential

The Schrödinger equation in D-dimensions is:

$$\frac{-\hbar^2}{2\mu} \left( \frac{d^2}{dr^2} + \frac{D-1}{r} \frac{d}{dr} - \frac{l(l+D-2)}{r^2} \right) R_{n,l}(r) + V(r) R_{n,l}(r) = E_{n,l} R_{n,l}(r) \quad D = 3N-3 \tag{11}$$

where $N$ is the number of particle and $l$ is the angular momenta [26].

Here, consider the Schrödinger equation for 6-body system with a potential $V(r)$ that depends only on the distance $r$ from the origin:

$$\frac{-\hbar^2}{2\mu} \left( \frac{d^2}{dr^2} + \frac{14}{r} \frac{d}{dr} - \frac{l(l+13)}{r^2} \right) R_{n,l}(r) + V(r) R_{n,l}(r) = E_{n,l} R_{n,l}(r) \tag{12}$$

$$\left( \frac{d^2}{dr^2} + \frac{14}{r} \frac{d}{dr} - \frac{l(l+13)}{r^2} \right) R_{n,l}(r) + \frac{2\mu}{\hbar^2} \left( E_{n,l} - V(r) \right) R_{n,l}(r) = 0 \tag{13}$$

By applying $U_{n,l} = R_{n,l} r^{\frac{D-1}{2}} = R_{n,l} r^7$, we can write:

$$\frac{dR_{n,l}}{dr} = \frac{dU_{n,l}}{dr} r^{-7} - 7 r^{-8} U_{n,l}, \qquad \frac{d^2 R_{n,l}}{dr^2} = \frac{d^2 U_{n,l}}{dr^2} r^{-7} - 14 r^{-8} \frac{dU_{n,l}}{dr} + 56 r^{-9} U_{n,l} \tag{14}$$

By substituting Eq. (14) in Eq. (13), we found the following form for Eq. (13):

$$\frac{d^2U_{n,l}}{dr^2}r^{-7} - 42r^{-9}U_{n,l} - l(l+13)r^{-9}U_{n,l} + \frac{2\mu}{\hbar^2}(E_{n,l} - V(r))U_{n,l}r^{-7} = 0 \tag{15}$$

By some summarizing, Eq. (15) changes to:

$$\frac{d^2U_{n,l}}{dr^2} + \frac{2\mu}{\hbar^2}\left(E_{n,l} - V(r) - \frac{\hbar^2(l+6)(l+7)}{2\mu r^2}\right)U_{n,l} = 0 \tag{16}$$

where $\mu$ is the reduced mass. It is suitable to introduce the Yukawa potential and use the Taylor expansion. So the potential takes the form:

$$V = -\frac{V_0 e^{-ar}}{r} = -V_0\frac{(1-ar)}{r} = -\frac{V_0}{r} + aV_0 \tag{17}$$

By putting Eq. (17) into Eq. (16), the Schrödinger equation changes to:

$$\frac{d^2U_{n,l}(r)}{dr^2} + \frac{2\mu}{\hbar^2}\left(E_{n,l} - aV_0 + \frac{V_0}{r} - \frac{\Omega}{r^2}\right)U_{n,l}(r) = 0 \tag{18}$$

where $\Omega$ is defined as $\Omega = \frac{\hbar^2(l+6)(l+7)}{2\mu}$. By choosing $\varepsilon_{n,l} = \frac{2\mu}{\hbar^2}(E_{n,l} - aV_0)$, $\beta = \frac{4\mu}{\hbar^2}V_0$ and $\gamma = \frac{2\mu}{\hbar^2}\Omega$, Eq. (18) changes to:

$$\frac{d^2U_{n,l}(r)}{dr^2} + \left(\varepsilon_{n,l} + \frac{\beta}{r} - \frac{\gamma}{r^2}\right)U_{n,l}(r) = 0 \tag{19}$$

In Supersymmetric Quantom Mechanics, the superpotential is defined as:

$$W_1 = -\frac{\hbar}{\sqrt{2\mu}}\left(A + \frac{B}{r}\right) \tag{20}$$

Substituting this superpotential into Riccati equation that has the form of:

$$W_1^2(x) - \frac{\hbar}{\sqrt{2\mu}}W_1'(x) = \frac{2\mu}{\hbar^2}\left(V_1(x) - E_0^{(1)}\right) \tag{21}$$

Then we reach to:

$$\left(A^2 + \frac{B^2}{r^2} + \frac{2AB}{r} - \frac{B}{r^2}\right) = \left(-\varepsilon_{n,l} - \frac{\beta}{r} + \frac{\gamma}{r^2}\right) \tag{22}$$

By doing some calculations, we can get $A^2 = -\varepsilon_{n,l}$, $2AB = -\beta$, $B^2 - B = \gamma$ and the ground state binding energy for Lithium nucleus is given as following:

$$E_{n,l} = \frac{-\hbar^2}{2\mu}\left(\frac{-\beta}{1\pm\sqrt{1+4\gamma}}\right)^2 + V_0 a \tag{23}$$

By using Eq. (24) from SUSY method:

$$\psi_0^1(r) = N_0 \exp\left(-\frac{\sqrt{2\mu}}{\hbar}\int W(r')\,dr'\right) \tag{24}$$

the ground state normalized eigenfunctions are given as:

$$\psi_0^1(r) = N_0 \exp[Ar + B\ln r] \tag{25}$$

In Table 1. the fitted values of parameters of the ground state binding energy equations for Yukawa potential are given.

Table 1. The fitted values of parameters of the ground state binding energy equations for Yukawa potential, coloumn B.E (our model) contains our calculation and the column B.E (experiment) contains the experimental data.

| $a$ (fm)$^{-1}$ | $V_0$ (MeV.fm) | B.E (MeV) | B.E (experiment) [27] |
|---|---|---|---|
| 0.59 | 50 | 29.39 | 31.995 |
| 0.70 | 45 | 31.42 | 31.995 |
| 0.95 | 33 | 31.31 | 31.995 |
| 0.70 | 40 | 27.93 | 31.995 |
| 0.80 | 40 | 31.93 | 31.995 |
| 0.80 | 30 | 23.96 | 31.995 |
| 0.60 | 50 | 29.90 | 31.995 |
| 0.50 | 50 | 24.90 | 31.995 |
| 0.90 | 35 | 31.45 | 31.995 |
| 0.64 | 50 | 31.90 | 31.995 |

From Table 1, it can be seen that for $a = 0.80$(fm$^{-1}$) and $V_0 = 40$(MeV), the calculated ground state binding energy (31.93MeV) has a good agreement with the experimental data.

As we know from SUSY, the potential is determined as:

$$V_\pm = W^2 \pm \frac{\hbar}{\sqrt{2\mu}}\frac{dW}{dr} = \frac{\hbar^2}{2\mu}\left[A^2 + \frac{B^2}{r^2} + \frac{2AB}{r^2} \pm \frac{B}{r^2}\right] \tag{26}$$

We obtain $A = \frac{-\beta}{2(B^2-\gamma)}$ from Eq. (22). By substituting $A$ in Eq. (26) we reach to:

$$V_\pm = \frac{\hbar^2}{2\mu}\left[(\frac{-\beta}{2(B^2-\gamma)})^2 + \frac{B^2}{r^2} + \frac{2\left(\frac{-\beta}{2(B^2-\gamma)}\right)B}{r^2} \pm \frac{B}{r^2}\right] \quad (27)$$

The shape invariance concept that was introduced by Gendenshtein is [26]:

$$V_+(a_0,r) = V_-(a_1,r) + R(a_1) \quad (28)$$

where $a_1$ is a function of $a_0$ and $R(a_1)$ is independent of $r$. Hence, the energy spectrum becomes:

$$E_0^{(k)} = \sum_{i=0}^{k} R(a_i) \quad , \quad E_{nl} = E_{nl}^- + E_0 \quad (29)$$

If we now consider a mapping of the form:

$$B \to B' = B - a \quad (30)$$

In Eq. (27), it is easily seen that apart from a constant, the partner potential are the same. In technical words, the chosen SUSY potential satisfies the shape invariance condition. On the other hand, we can obtain:

$$B_1 = B_0 - a \quad , \quad B_n = B_0 - na \quad (31)$$

$$R(a_1) = V_+(B,r) - V_-(B-a,r) =$$
$$\frac{\hbar^2}{2\mu}\left[(\frac{-\beta}{2(B^2-\gamma)})^2 + \frac{B^2}{r^2} + \frac{2\left(\frac{-\beta}{2(B^2-\gamma)}\right)B}{r^2} + \frac{B}{r^2} - (\frac{-\beta}{2((B-a)^2-\gamma)})^2 - \frac{(B-a)^2}{r^2} - \frac{2\left(\frac{-\beta}{2((B-a)^2-\gamma)}\right)(B-a)}{r^2} + \frac{B-a}{r^2}\right] \quad (32)$$

$$\to R(a_i) = -\frac{\hbar^2}{2\mu}\left[(\frac{-\beta}{2((B-ia)^2-\gamma)})^2 - (\frac{-\beta}{2((B-(i-1))^2-\gamma)})^2\right] \quad (33)$$

$$R(a_i) = -\frac{\hbar^2}{2\mu}\left[(\frac{-\beta}{2((B-ia)^2-\gamma)})^2 - (\frac{-\beta}{2((B-(i-1)a)^2-\gamma)})^2\right] \quad (34)$$

Where the remainder $R(a_i)$ is independent of $r$. By using Eqs. (23) and (29), the energy levels of the Yukawa potential are found as:

$$E_{n,l} = -\frac{\hbar^2}{2\mu}\left[(\frac{-\beta}{2((B-na)^2-\gamma)})^2 - (\frac{-\beta}{2(B^2-\gamma)})^2 + \left(\frac{-\beta}{1+\sqrt{1+4\gamma}}\right)^2\right] + aV_0 \quad (35)$$

Now, we change the potential in Eq. (16). For getting better results, we use the generalized Yukawa potential that takes the form of:

$$V = \frac{-V_0 e^{-ar}}{r} + \frac{k}{r^2} \qquad (36)$$

By substituting Eq. (36) in Eq. (16), we reach to:

$$\frac{d^2 U_{n,l}(r)}{dr^2} + \frac{2\mu}{\hbar^2}\left(E_{n,l} + \frac{V_0 e^{-ar}}{r} - \frac{k}{r^2} - \frac{\hbar^2(l+6)(l+7)}{2\mu r^2}\right) U_{n,l}(r) = 0 \qquad (37)$$

By using Taylor expansion, Eq. (37) can be written:

$$\frac{d^2 U_{n,l}(r)}{dr^2} + \frac{2\mu}{\hbar^2}\left(E_{n,l} - aV_0 + \frac{V_0}{r} - \frac{\Omega + k}{r^2}\right) U_{n,l}(r) = 0 \qquad (38)$$

where $\Omega$ is defined as $\Omega = \frac{\hbar^2 (l+6)(l+7)}{2\mu}$. If we get $\varepsilon_{n,l} = \frac{2\mu}{\hbar^2}(E - aV_0)$, $\beta = \frac{4\mu}{\hbar^2} V_0$ and $\gamma = \frac{2\mu}{\hbar^2}(\Omega + k)$, then we reach to:

$$\frac{d^2 U_{n,l}(r)}{dr^2} + \left(\varepsilon_{n,l} + \frac{\beta}{r} - \frac{\gamma}{r^2}\right) U_{n,l}(r) = 0 \qquad (39)$$

As we know from SUSY the superpotential define as Eq. (20). When we use this into Ricatti equation (Eq. (21)), as we do before, gives us $A^2 = -\varepsilon_{n,l}$, $2AB = -\beta$ and $B^2 - B = \gamma$. So the ground state binding Energy is obtained as:

$$E_{n,l} = \frac{-\hbar^2}{2\mu}\left(\frac{-\beta}{1 \pm \sqrt{1 + 4\gamma}}\right)^2 + V_0 a \qquad (40)$$

by using Eq. (24) from SUSY method, the ground state normalized eigenfunctions are given as Eq. (25). In Table 2. the fitted values of parameters of the ground state binding energy equations for the generalized Yukawa potential are given.

From Table 2, it can be seen that for $a = 0.80 \text{(fm}^{-1})$, $k=30$(MeV) and $V_0=40$(MeV), the calculated ground state binding energy (31.98 MeV) has a good agreement with the experimental data.

Table 2. The fitted values of parameters of the ground state binding energy equations for the generalized Yukawa potential, column B.E (our model) contains our calculation and the column B.E (experiment) contains the experimental data.

| $a$ (fm)$^{-1}$ | $V_0$ (MeV.fm) | B.E (MeV) | B.E (experiment) [27] | $k$ (MeV.fm$^2$) |
|---|---|---|---|---|
| 0.70 | 45 | 31.42 | 31.995 | 5 |
| 0.60 | 50 | 29.90 | 31.995 | 10 |
| 0.64 | 49.3 | 31.90 | 31.995 | 50 |
| 0.70 | 50 | 34.90 | 31.995 | 30 |
| 0.75 | 40 | 29.93 | 31.995 | 30 |
| 0.50 | 50 | 24.90 | 31.995 | 30 |
| 0.55 | 50 | 27.40 | 31.995 | 50 |
| 0.80 | 30 | 23.96 | 31.995 | 50 |
| 0.80 | 40 | 31.98 | 31.995 | 30 |
| 0.59 | 50 | 29.40 | 31.995 | 50 |

As we did before, the energy levels with this potential take the form of:

$$E_{n,l} = -\frac{\hbar^2}{2\mu}\left[\left(\frac{-\beta}{2((B-na)^2-\gamma)}\right)^2 - \left(\frac{-\beta}{2(B^2-\gamma)}\right)^2 + \left(\frac{-\beta}{1+\sqrt{1+4\gamma}}\right)^2\right] + V_0 a \qquad (41)$$

Another potential that we use in Eq. (16) is Hellmann potential. So the Schrödinger equation takes the form of:

$$\frac{d^2 U_{n,l}(r)}{dr^2} + \frac{2\mu}{\hbar^2}\left(E_{n,l} + V_0 a + \frac{k-V_0}{r} - \frac{\Omega}{r^2}\right)U_{n,l}(r) = 0 \qquad (42)$$

where $\Omega$ is defined as $\Omega = \frac{\hbar^2(l+6)(l+7)}{2\mu}$. By choosing $\varepsilon_{n,l} = \frac{2\mu}{\hbar^2}(E_{n,l} + aV_0)$, $\beta = \frac{2\mu}{\hbar^2}(k-V_0)$ and $\gamma = \frac{2\mu}{\hbar^2}\Omega$, Eq. (41) takes the form of:

$$\frac{d^2 U_{n,l}(r)}{dr^2} + \left(\varepsilon_{n,l} + \frac{\beta}{r} - \frac{\gamma}{r^2}\right)U_{n,l}(r) = 0 \qquad (43)$$

We define the suprepotential as Eq. (23) and do the same for this potential we can obtain the Binding energy in the form of

$$E_{n,l} = \frac{-\hbar^2}{2\mu}\left(\frac{-\beta}{1\pm\sqrt{1+4\gamma}}\right)^2 - V_0 a \qquad (44)$$

By using Eq. (24) from SUSY method, the ground state normalized eigenfunctions are given as Eq. (25).

In Table 3. the fitted values of parameters of the ground state binding energy equations for Hellmann potential are given.

As what we did before for obtaining $E_{n,l}$ do here and it takes the form of

$$E_{n,l} = -\frac{\hbar^2}{2\mu}\left[\left(\frac{-\beta}{2((B-na)^2-\gamma)}\right)^2 - \left(\frac{-\beta}{2(B^2-\gamma)}\right)^2 + \left(\frac{-\beta}{1+\sqrt{1+4\gamma}}\right)^2\right] - V_0 a \quad (45)$$

Table 3. The fitted values of parameters of the ground state binding energy equations for Hellmann potential, column B.E (our model) contains our calculation and the column B.E (experiment) contains the experimental data.

| $a$ (fm)$^{-1}$ | $V_0$ (MeV.fm) | B.E (MeV) | B.E (experiment) [27] | $k$ (MeV.fm) |
|---|---|---|---|---|
| 0.60 | -50 | 29.85 | 31.995 | 10 |
| 0.60 | -50 | 29.80 | 31.995 | 20 |
| 0.65 | -50 | 32.38 | 31.995 | 5 |
| 0.58 | -50 | 28.85 | 31.995 | 10 |
| 0.65 | -49 | 31.70 | 31.995 | 10 |
| 0.65 | -49 | 31.65 | 31.995 | 20 |
| 0.70 | -40 | 27.85 | 31.995 | 20 |
| 0.71 | -45 | 31.83 | 31.995 | 10 |
| 0.50 | -50 | 24.85 | 31.995 | 10 |
| 0.55 | -50 | 27.35 | 31.995 | 10 |

From Table 3, it can be seen that for $a = 0.71$(fm$^{-1}$), $k=10$(MeV) and $V_0=-45$(MeV), the calculated ground state binding energy (31.83 MeV) has a good agreement with the experimental data.

## 6. Conclusion

In this paper, we have obtained the exact solution of the Schrödinger equation for Yukawa potential, generalized Yukawa potential and Hellmann potential within the frame work of SUSYQM. Also, we have calculated the binding energy of lithium nucleus for ground state with various potentials. Then, we have found the wave function for this element. The results obtained from SUSY method for these three potentials are in a good agreement with the experimental data.